\documentclass[a4paper,12pt]{article}
\usepackage{float}
\usepackage{tabularx}
\usepackage{amsmath}
\usepackage{amsfonts}
\usepackage{braket}
\usepackage{adjustbox}
\usepackage{multirow}
\usepackage{color}
\usepackage{longtable}
\usepackage{pdflscape}
\usepackage{rotating}
\usepackage{xcolor}
\usepackage[colorlinks = true,
            linkcolor = blue,
            urlcolor  = blue,
            citecolor = blue,
            anchorcolor = blue]{hyperref}
\usepackage[margin=1in,footskip=0.3in]{geometry}
\title{Exact Neutron-Proton Wavefunctions Using the Phase Function Method}

\author{Anil Khachi$^{*}$\\\\
$^{*}$Chandigarh Group of Colleges Jhanjeri, Mohali, Punjab, India- 140307\\  Chandigarh Engineering College, Department of Applied Sciences}
\begin{document}
\maketitle
\abstract{\noindent 
Radial phase shifts ($\delta(r)$), amplitude functions ($A(r)$), and exact wavefunctions ($u(r)$) for various uncoupled S, P, and D channels of neutron--proton scattering have been calculated using the Phase Function Method (PFM). In these calculations, inverse potentials obtained from the Morse function as the zeroth-order reference potential are employed. The parameters of the Morse potential were optimized using the comprehensive GRANADA partial wave analysis, consisting of 6713 experimental \textit{np} phase shift data points from 1950 to 2013, by minimizing the mean square error (MSE) as a cost function. The present work provides detailed radial dependence of $\delta(r)$, $A(r)$, and $u(r)$ up to 5~fm for laboratory energies $E_{\ell \text{lab}} = [1, 10, 50, 100, 150, 250, 350]$~MeV. The obtained wavefunctions show excellent agreement with high-precision Nijmegen-II results, highlighting the accuracy and transparency of the PFM approach for uncoupled scattering states.
}\\
{\textbf{keywords:} n-p scattering, phase function method (PFM), Morse potential, exact wavefunctions, uncoupled states}

\section{Introduction}
Under ``standard approach" main objective of theoretical physicst is to calculate the
wavefunction from which other scattering properties like phase shifts, low energy scattering
parameters, static properties etc., can be obtained. It is to be noted that the experimentalists do not get wavefunction as an output of an experiment, rather one gets outputs like differential cross sections and total cross-sections, transition energies etc., which are a result of the interaction processes. Phase shifts are further related to the cross sections. The interaction is portrayed from the experimental outputs in form of V(r) vs r (fm) plots. According to Schwinger, Landau and Smorodinsky \cite{1} ``The meeting place between theory and experiment is not the phase shifts themselves but the value of the variational parameters implied by phase shifts".
\newline There are various techniques that can be used to obtain the scattering phases by solving the
Schrödinger equation like: Born approximation, Brysk's approximation,and other successive approximation techniques. In our earlier papers we have applied PFM or VPA for studying n-n \cite{nn}, n-p \cite{2,3,4}, p-p \cite{pp}, n-d \cite{5}, $\alpha-\alpha$ \cite{alp} n-$\alpha$ \cite{nalp}, $p-D$ \cite{pd} and $\alpha-(^{3}H ~\&~ ^{3}He)$ using Morse and Malfliet-Tjon potential \cite{MT}. Knowing that using PFM, phase
shifts of the nucleon-nucleon scattering can be obtained using different types of potentials (high
precision phenomenological potentials) and interaction models between two nucleons. This paper is an extension of our recent published work \cite{8} where we have considered Morse potential \cite{7} as nuclear interaction for obtaining scattering phase shifts (SPS) for various $\ell$ channels in \textit{np} scattering. The current paper deals with
using the obtained parameters \cite{8} and using those parameters we have obtained various functions like $\delta$ vs r,$A$ vs r and $u$ vs r.

\section{Methodology}
The Morse function is given by \cite{7}
\begin{equation}
V(r) = V_0 (e^{-2 \frac{(r-r_m)}{a_m}}-2 e^{-\frac{(r-r_m)}{a_m}})
\end{equation}
In above equation, the parameters $V_0$, $r_m$, and $a_m$ represent the strength of interaction between the particles, the equilibrium distance at which maximum attraction is felt, and the shape of the potential, respectively. The Morse potential possesses several distinguishing properties that differentiate it from other phenomenological potentials, including:
(i) The Schr$\ddot{\text{o}}$dinger equation is exactly solvable for this potential.(ii) Unlike other phenomenological potentials such as Manning-Rosen \cite{9}, Malfliet-Tjon \cite{10}, Hulthén \cite{11}, and others, the Morse potential has an exact analytical expression for the $^1S_0$ state.(iii) It exhibits a relatively simpler wavefunction.(iv) It is a shape-invariant potential.
Hence, it can be considered as a good choice for modeling the interaction between any two scattering particles.\\
\subsection{Phase Function Method}  
The Schr$\ddot{o}$dinger wave equation for a spinless particle with energy E and orbital angular momentum $\ell$ undergoing scattering is given by
\begin{equation}
\frac{\hbar^2}{2\mu} \bigg[\frac{d^2}{dr^2}+\big(k^2-\ell(\ell+1)/r^2\big)\bigg]u_{\ell}(k,r)=V(r)u_{\ell}(k,r)
\label{Scheq}
\end{equation}
Where $k=\sqrt{E/(\hbar^2/2\mu)}$.

Second order differential equation  Eq.\ref{Scheq} has been transformed to the first order non-homogeneous differential equation of Riccati type \cite{12,13} given by
\begin{equation}
\delta_{\ell}'(k,r)=-\frac{V(r)}{{k\left(\hbar^{2} / 2 \mu\right)}}\bigg[\cos(\delta_\ell(k,r))\hat{j}_{\ell}(kr)-\sin(\delta_\ell(k,r))\hat{\eta}_{\ell}(kr)\bigg]^2
\label{PFMeqn}
\end{equation}
 
Prime denotes differentiation of phase shift with respect to distance and the Riccati Hankel function of first kind is related to $\hat{j_{\ell}}(kr)$ and $\hat{\eta_{\ell}}(kr)$ by $\hat{h}_{\ell}(r)=-\hat{\eta}_{\ell}(r)+\textit{i}~ \hat{j}_{\ell}(r)$ . In integral form the above equation can be written as
\begin{equation}
\delta(k,r)=\frac{-1}{{k\left(\hbar^{2} / 2 \mu\right)}}\int_{0}^{r}{V(r)}\bigg[\cos(\delta_{\ell}(k,r))\hat{j_{\ell}}(kr)-\sin(\delta_{\ell}(k,r))\hat{\eta_{\ell}}(kr)\bigg]^2 dr
\end{equation}


Eq.\ref{PFMeqn} is numerically solved using Runge-Kutta 5$^{th}$ order (RK-5) method \cite{14} with initial condition $\delta_{\ell}(0) = 0$. For $\ell = 0$, the Riccati-Bessel and Riccati-Neumann functions $\hat{j}_0$ and $\hat{\eta}_0$ get simplified as $\sin(kr)$ and $-\cos(kr)$, so Eq.\ref{PFMeqn}, for $\ell = 0$ becomes 
\begin{equation}
\delta'_0(k,r)=-\frac{V(r)}{{k\left(\hbar^{2} / 2 \mu\right)}}\sin^2[kr+\delta_0(k,r)]
\end{equation}
In above equations $k=\sqrt{E/(\hbar^2/2\mu)}$ and $\hbar^2/2\mu$ = 41.47 MeV fm$^{2}$ for \textit{np} case.
In above equation the function $\delta_0(k,r)$ was termed ``Phase function'' by Morse and Allis \cite{15}.
Similarly by varying the Bessel functions for various l values by using following recurrence relations \cite{16}
\begin{equation} 
\hat{j}_{{\ell}+1}(k r)=\frac{2 {\ell}+1}{k r} \hat{j}_{\ell}(k r)-\hat{j}_{{\ell}-1}(k r)\\ 
\end{equation}\begin{equation}
    \hat{\eta}_{{\ell}+1}(k r)=\frac{2 {\ell}+1}{k r} \hat{\eta}_{\ell}(k r)-\hat{\eta}_{{\ell}-1}(k r)
\end{equation}\newline  we obtain PFM equation for P-wave having following form \\
 \begin{equation}
    \delta_{1}^{\prime}(k, r)=\frac{-V(r)}{k\left(\hbar^{2} / 2 \mu\right)}\left[\frac{\sin \left(\delta_{1}+k r\right)-k r \cdot \cos \left(\delta_{1}+k r\right)}{k r}\right]^{2}
\end{equation}\newline PFM equation for D-wave takes following form\\
 \begin{equation}
    \delta_{2}^{\prime}(k, r)=\frac{-V(r)}{k\left(\hbar^{2} / 2 \mu\right)})\left[-\sin \left(k r+\delta_{2}\right)-3 \cos \frac{\left(\delta_{2}+k r\right)}{k r}+3 \sin \frac{\left(\delta_{2}+k r\right)}{(k r)^{2}}\right]^{2} 
 \end{equation}
The equation for  amplitude function\cite{17} with initial condition is obtained in the form

\begin{equation}
\begin{aligned}
    A_{\ell}^{\prime}(r) = &-\frac{A_{\ell} V(r)}{k} \left[\cos (\delta_\ell(k,r)) \hat{j}_{\ell}(kr)-\sin (\delta_\ell(k,r)) \hat{\eta}_{\ell}(kr)\right] \\
    &\times\left[\sin (\delta_\ell(k,r))( \hat{j}_{\ell}(kr)+\cos (\delta_\ell(k,r)) \hat{\eta}_{\ell}(kr)\right]
\end{aligned}
\end{equation}
\newline
also the equation to obtained wavefunction\cite{17} is 
\begin{equation}
    u_{\ell}(r)=A_{\ell}(r)\left[\cos (\delta_\ell(k,r)) \hat{j}_{\ell}(k r)-\sin (\delta_\ell(k,r)) \hat{\eta}_{\ell}(k r)\right]
\end{equation}
\begin{figure}[H]
    \centering
    \includegraphics[width=1.05\linewidth]{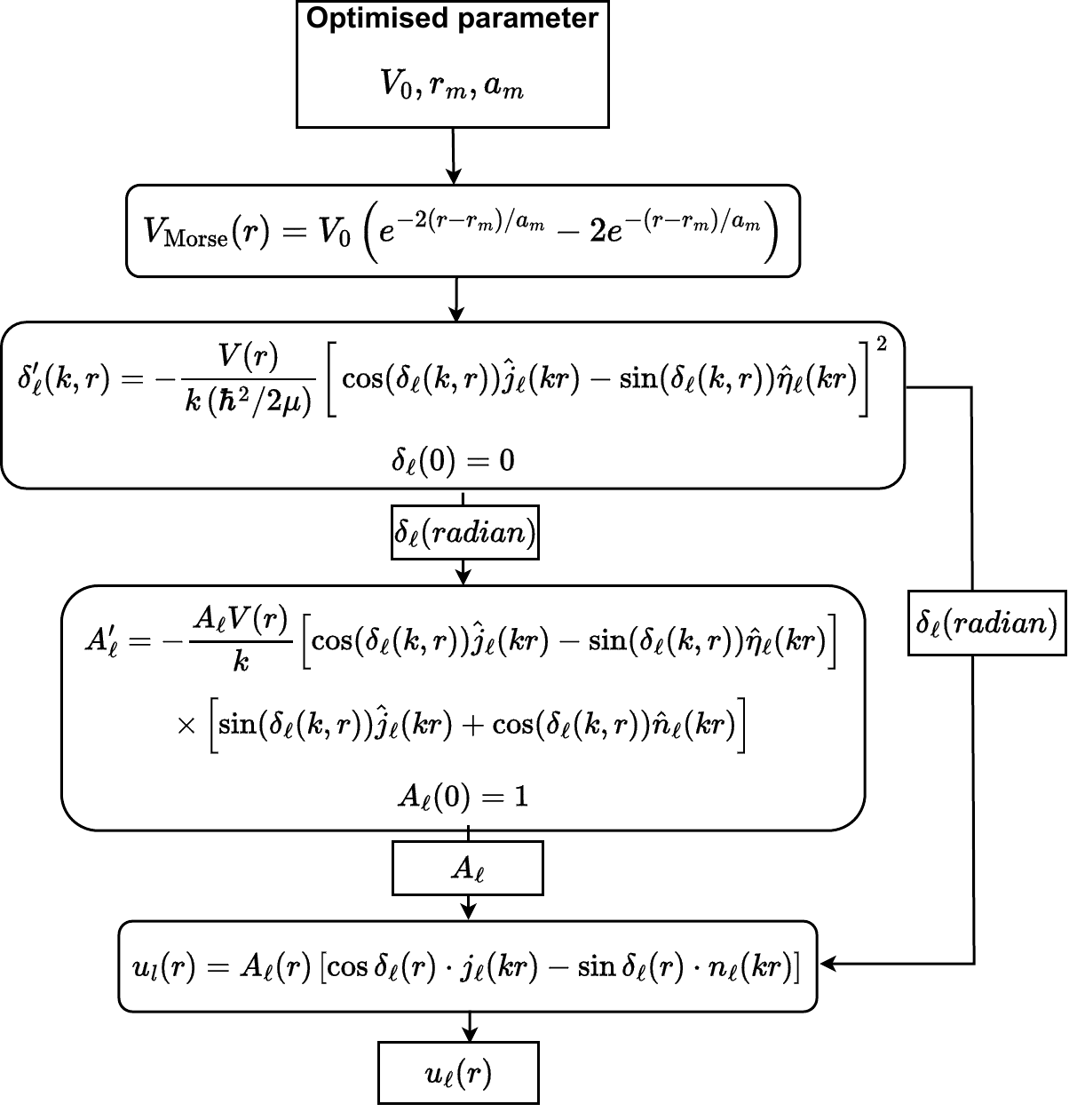}
    \caption{Detailed flowchart for obtaining phase shift, amplitude function and wavefunction.}
    \label{}
\end{figure}
\newpage
\section{Results and Discussion}
\begin{table}[ht!]
\centering
\caption{Model Parameters for Morse potential for various channels of n-p scattering \cite{8}.}
\scalebox{1.0}{
\begin{tabular}{cccc|c}

\hline\hline
States      & $V_0 (MeV)$    & $r_m(fm)$  & $a_m(fm)$   & $MSE$  \\ 
\hline
 $^{1}S_{0}$ & 70.438 & 0.901 &0.372 &0.649\\ 

  $^{1}P_{1}$ &0.010 & 5.442 &1.016 &1.568 \\

  $^{3}P_{0}$  & 11.579 & 1.750 &0.601 &  0.049 \\
 
  $^{3}P_{1}$  & 0.010 & 4.514 &0.778 &  0.832 \\
 
  $^{1}D_{2}$ & 131.302 & 0.010 &0.526 &0.026\\
  
  $^{3}D_{2}$ & 106.379 & 0.209 &0.747 &0.066\\
  
  %
  %
  %
  \hline
\end{tabular}

}
\end{table}

 \begin{figure}
 \centering
    \includegraphics[width=0.8\linewidth]{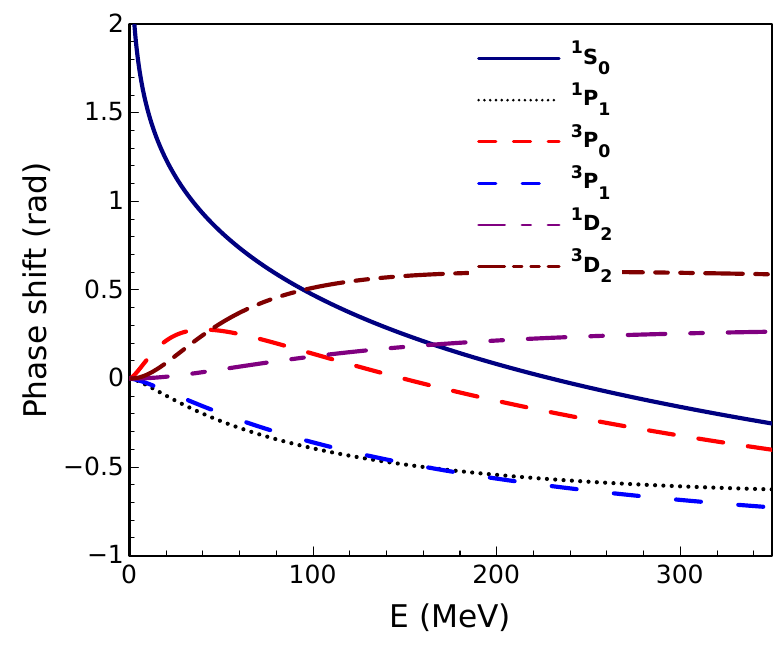}
    \includegraphics[width=0.8\linewidth]{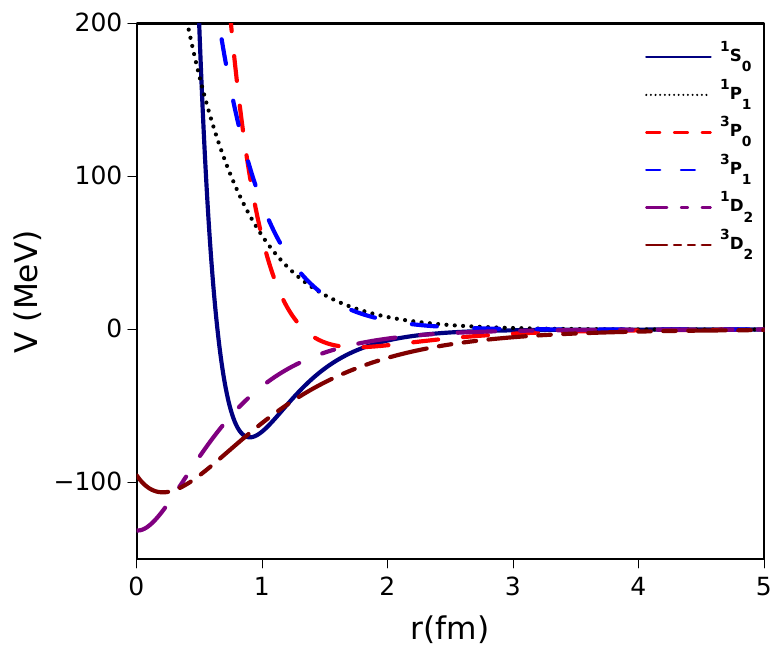}
    \caption{(top)Phase shifts $\delta(E)$ for various uncoupled scattering states are shown with respect to energy $E(MeV)$. (bottom) Respective interaction potentials for various scattering states.}
    \label{sps_pot}
\end{figure}

 \begin{figure}
    \includegraphics[width=0.55\linewidth]{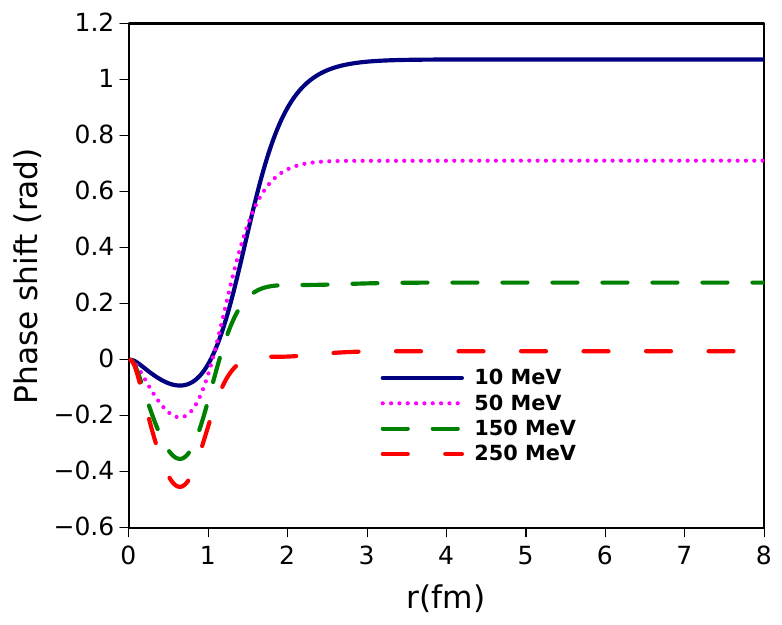}
    \includegraphics[width=0.55\linewidth]{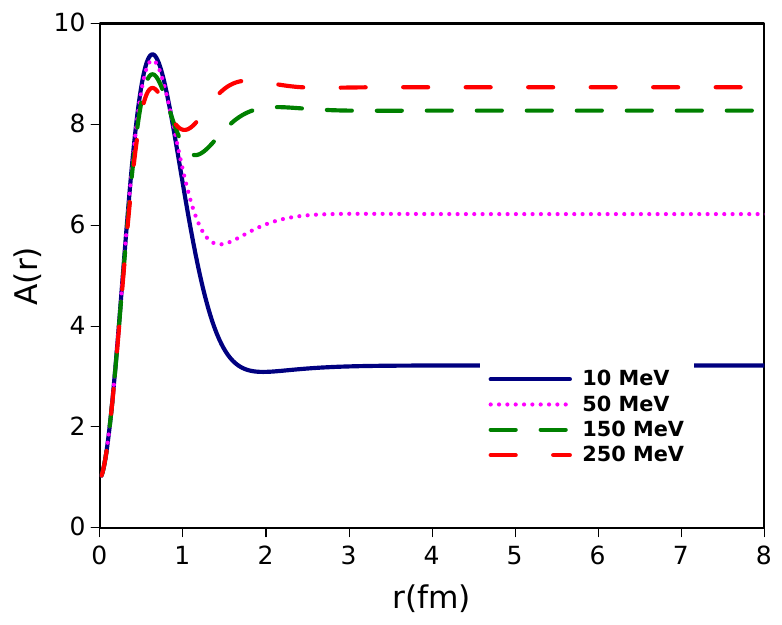}
    \includegraphics[width=0.55\linewidth]{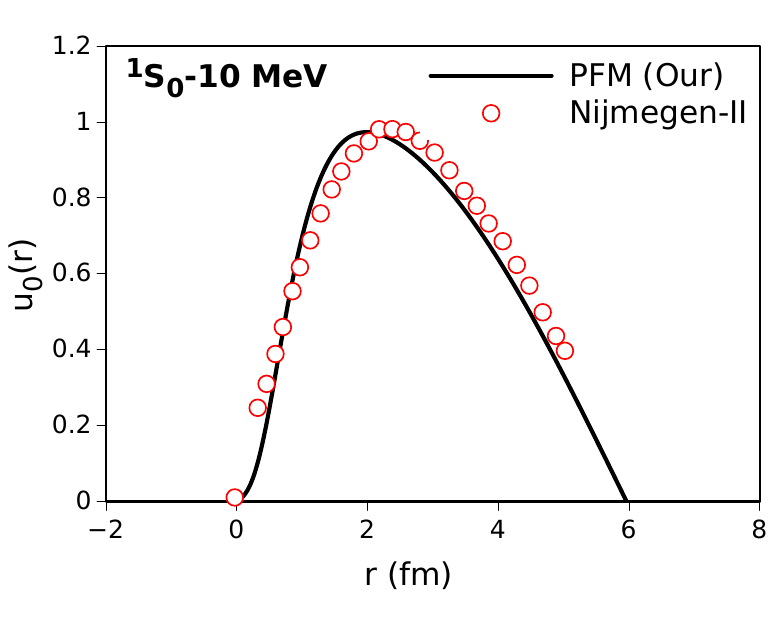}
    \includegraphics[width=0.55\linewidth]{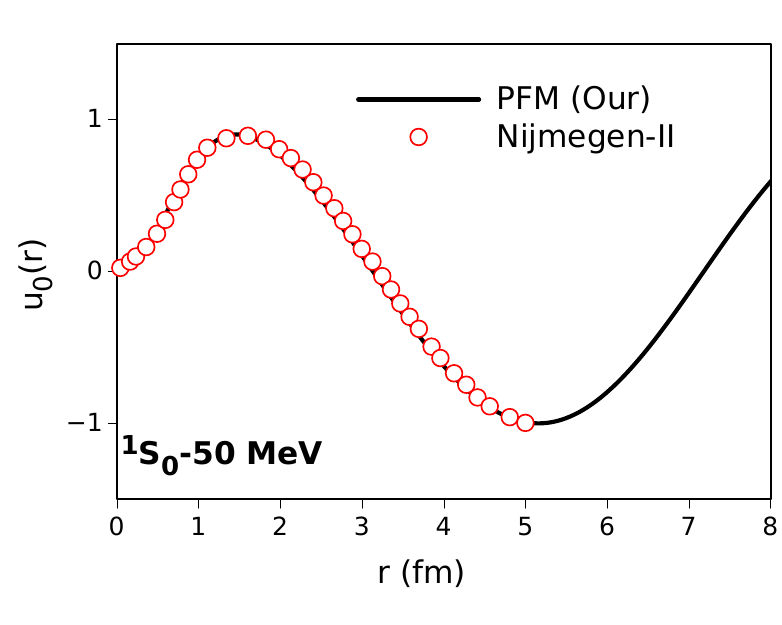}
    \includegraphics[width=0.55\linewidth]{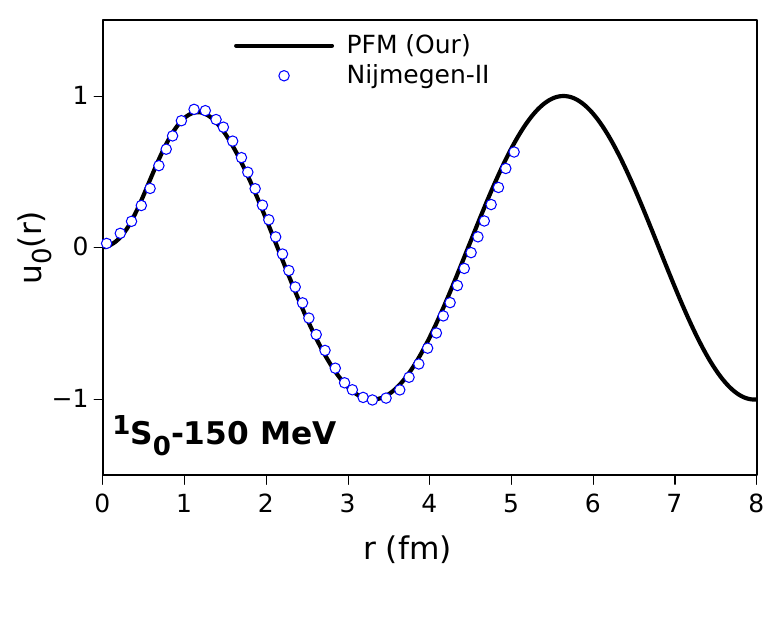}
    \includegraphics[width=0.55\linewidth]{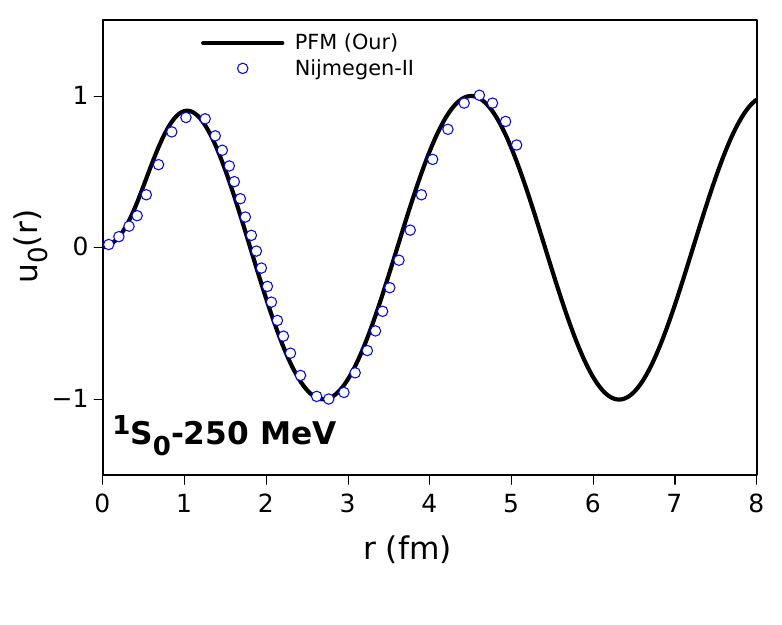}
    \caption{
Phase function, amplitude function, and radial wavefunctions for the $^{1}S_{0}$ channel.
Panel (a) shows the variation of the phase shift $\delta_{0}(r)$, panel (b) the amplitude
function $A_{0}(r)$, and panels (c--f) the corresponding wavefunctions $u_{0}(r)$ at
$E_{\text{lab}} = 10,\; 50,\; 150,$ and $250$~MeV. Results obtained using PFM are compared
with Nijmegen-II calculations.
}

    \label{1S0}
\end{figure}

 \begin{figure}
      \includegraphics[width=0.55\linewidth]{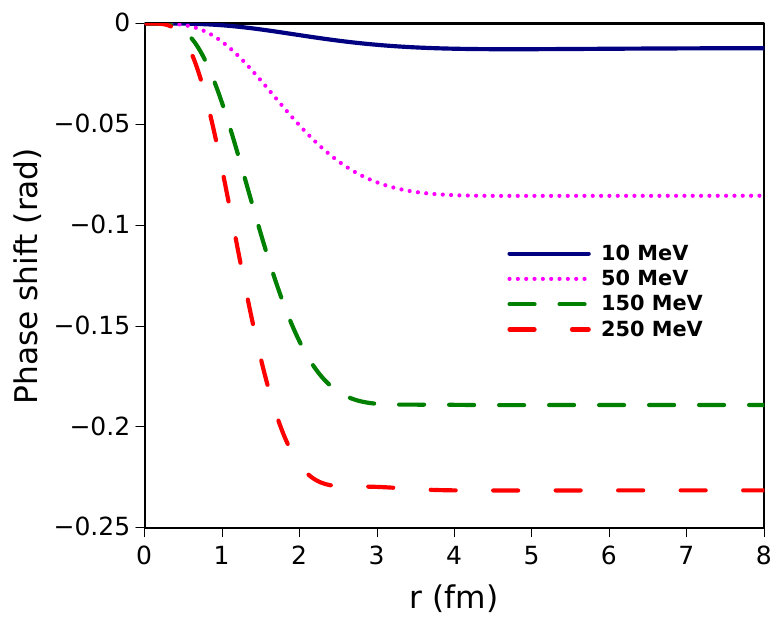}
      \includegraphics[width=0.55\linewidth]{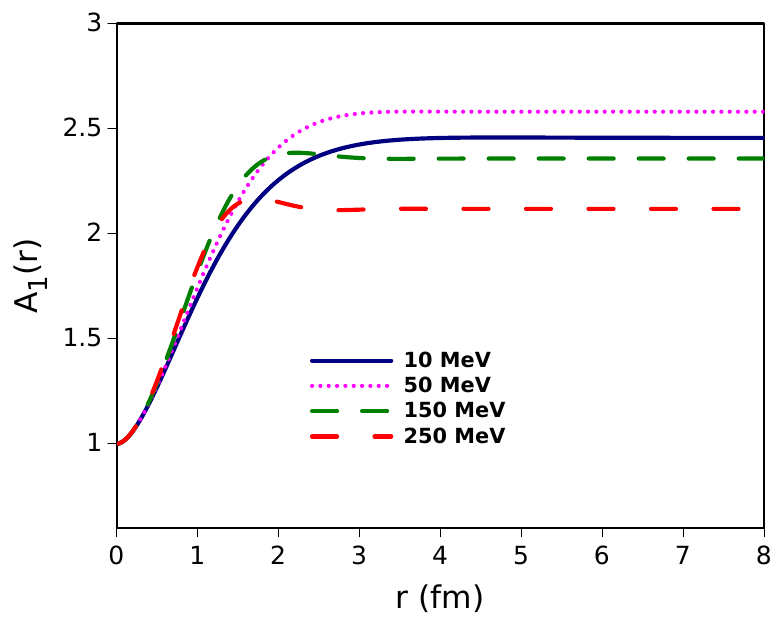}
    \includegraphics[width=0.55\linewidth]{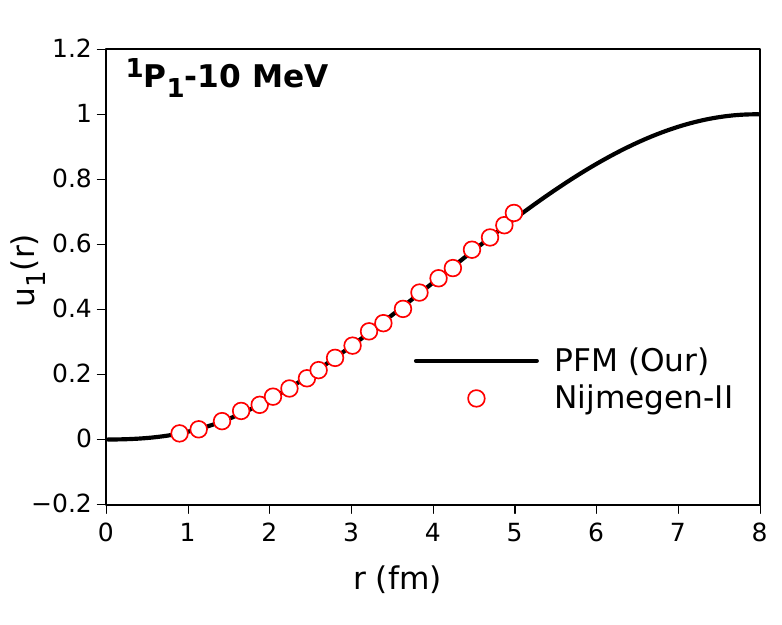}
    \includegraphics[width=0.55\linewidth]{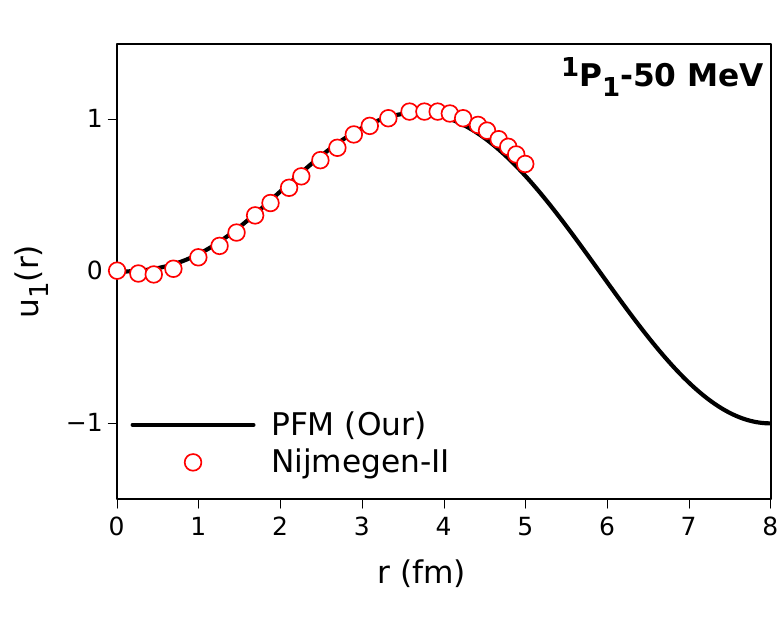}
    \includegraphics[width=0.55\linewidth]{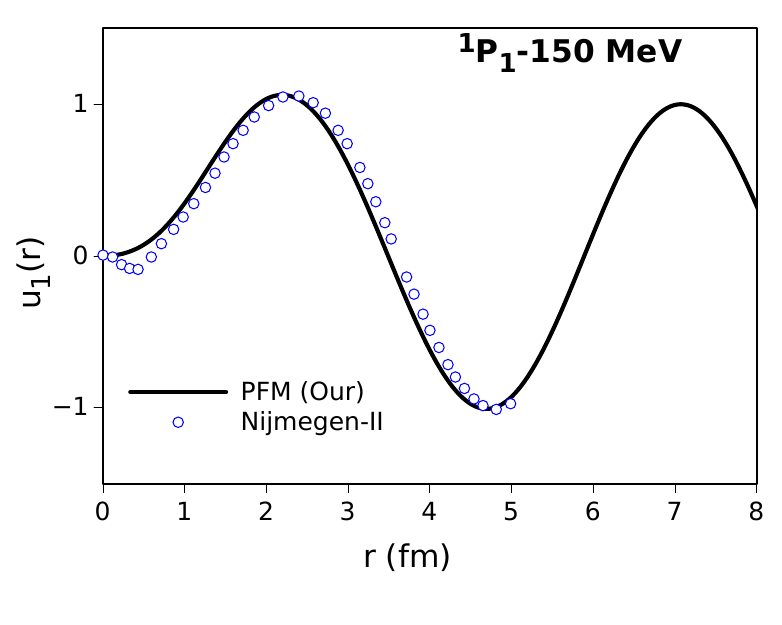}
    \includegraphics[width=0.55\linewidth]{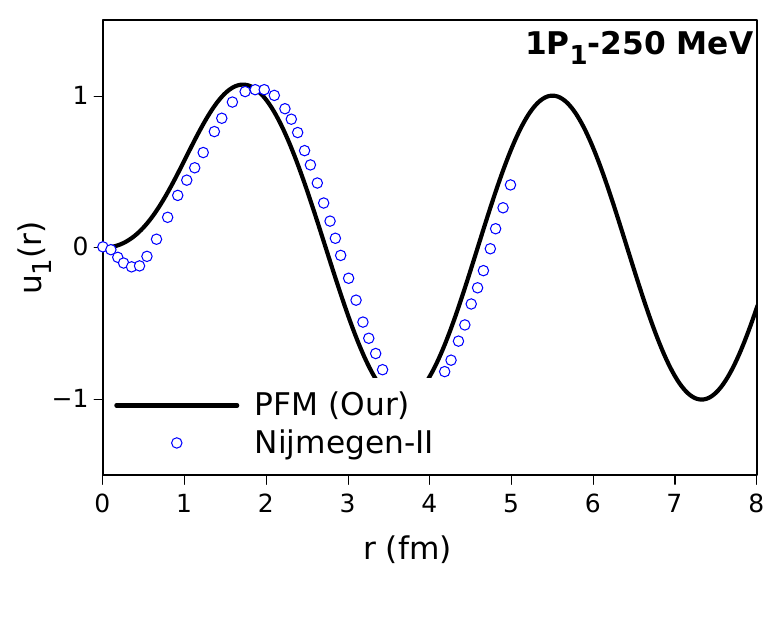}
    \caption{
Phase function, amplitude function, and radial wavefunctions for the $^{1}P_{1}$ channel.
Panel (a) shows the variation of the phase shift $\delta_{0}(r)$, panel (b) the amplitude
function $A_{0}(r)$, and panels (c--f) the corresponding wavefunctions $u_{0}(r)$ at
$E_{\text{lab}} = 10,\; 50,\; 150,$ and $250$~MeV. Results obtained using PFM are compared
with Nijmegen-II calculations.
}

    \label{fig:3D2}
\end{figure} 

 \begin{figure}
      \includegraphics[width=0.55\linewidth]{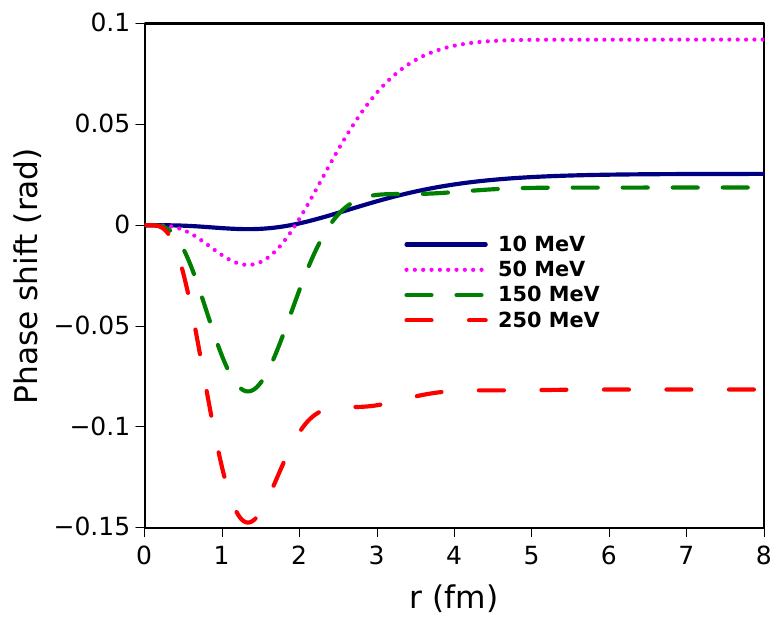}
      \includegraphics[width=0.55\linewidth]{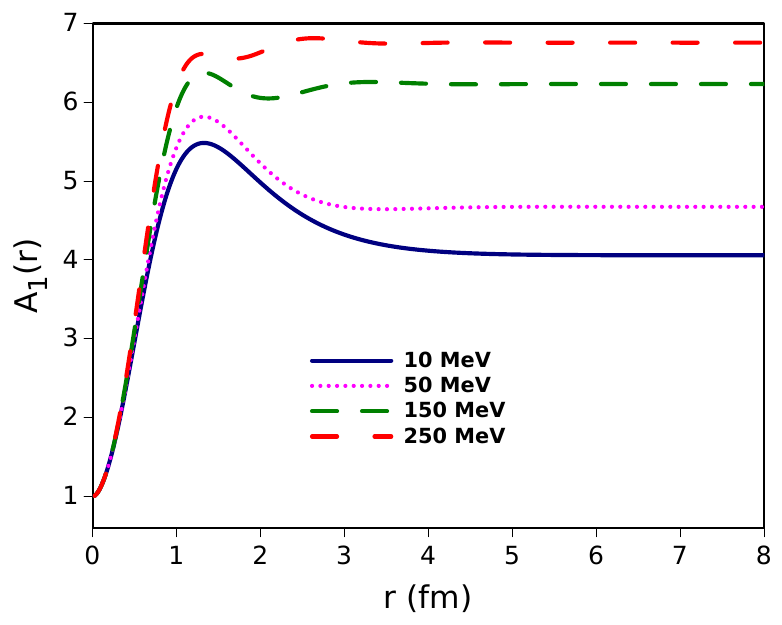}
    \includegraphics[width=0.55\linewidth]{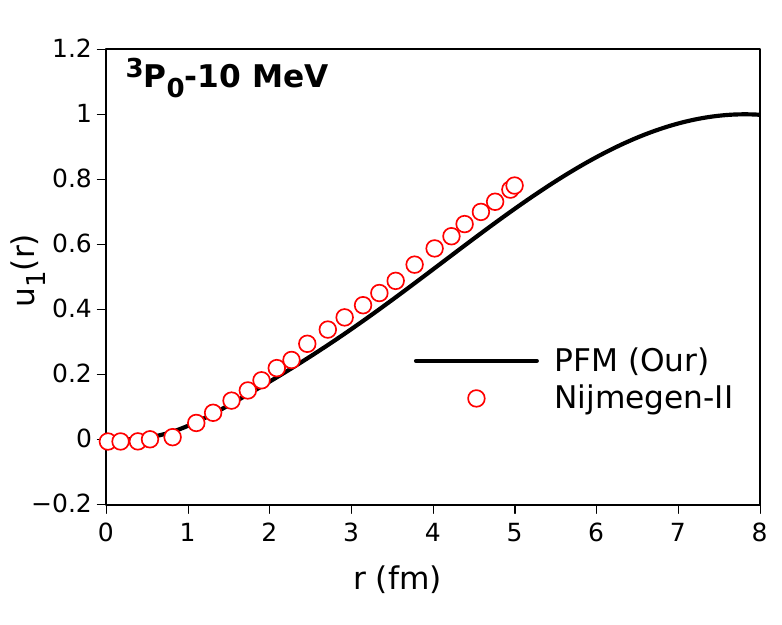}
    \includegraphics[width=0.55\linewidth]{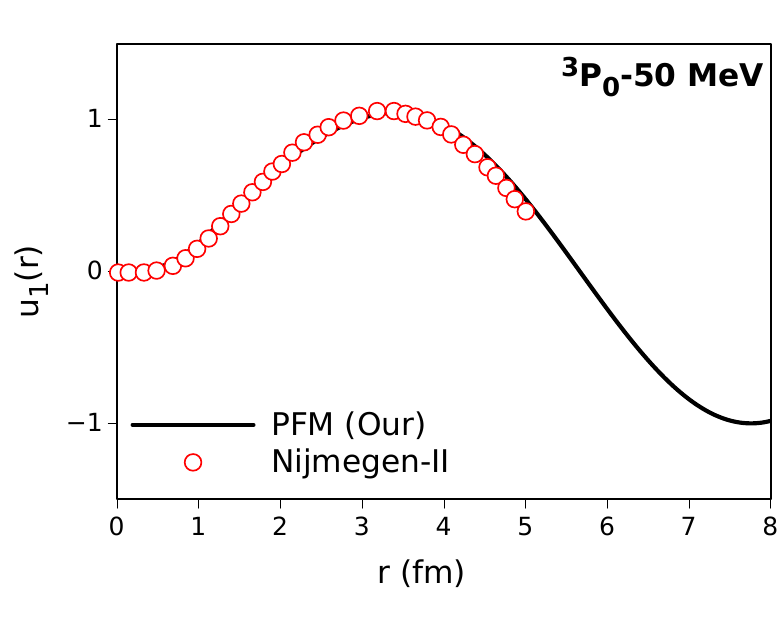}
    \includegraphics[width=0.55\linewidth]{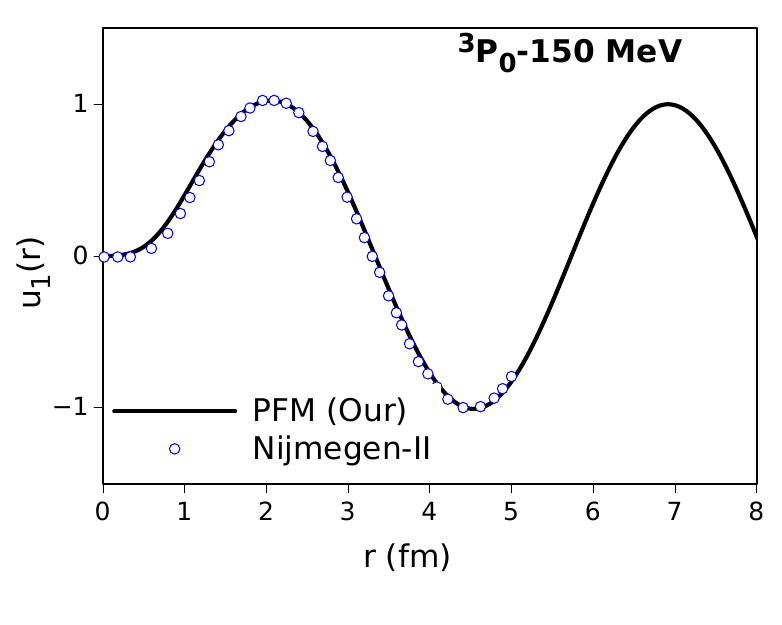}
    \includegraphics[width=0.55\linewidth]{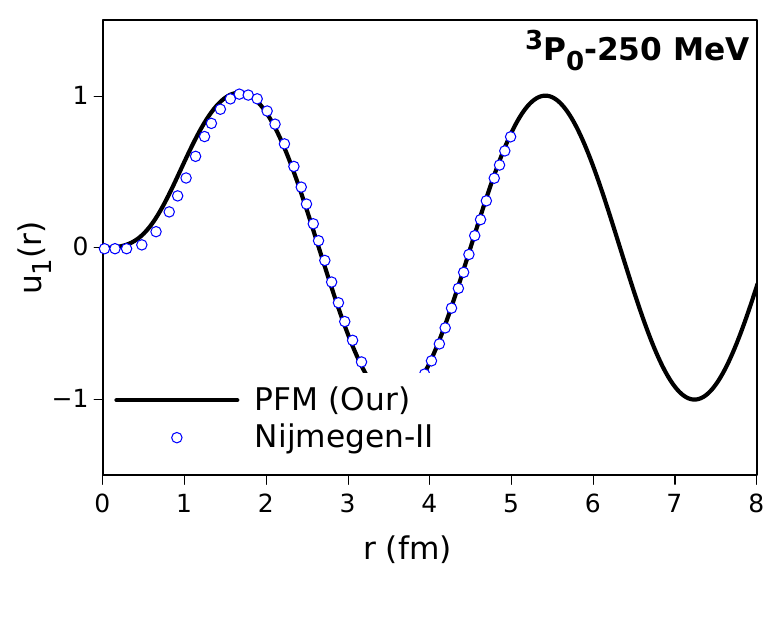}
    \caption{
Phase function, amplitude function, and radial wavefunctions for the $^{3}P_{0}$ channel.
Panel (a) shows the variation of the phase shift $\delta_{0}(r)$, panel (b) the amplitude
function $A_{0}(r)$, and panels (c--f) the corresponding wavefunctions $u_{0}(r)$ at
$E_{\text{lab}} = 10,\; 50,\; 150,$ and $250$~MeV. Results obtained using PFM are compared
with Nijmegen-II calculations.
}

    \label{fig:3D2}
\end{figure} 

 \begin{figure}
      \includegraphics[width=0.55\linewidth]{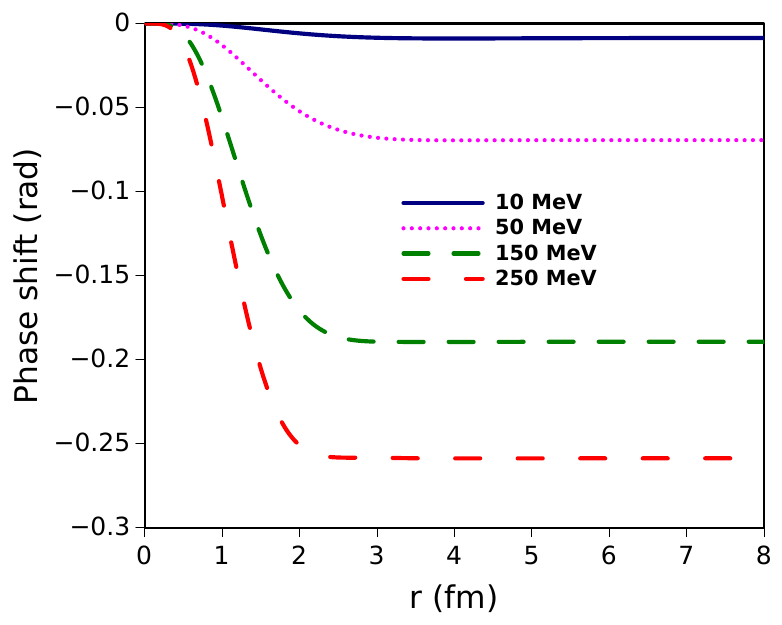}
      \includegraphics[width=0.55\linewidth]{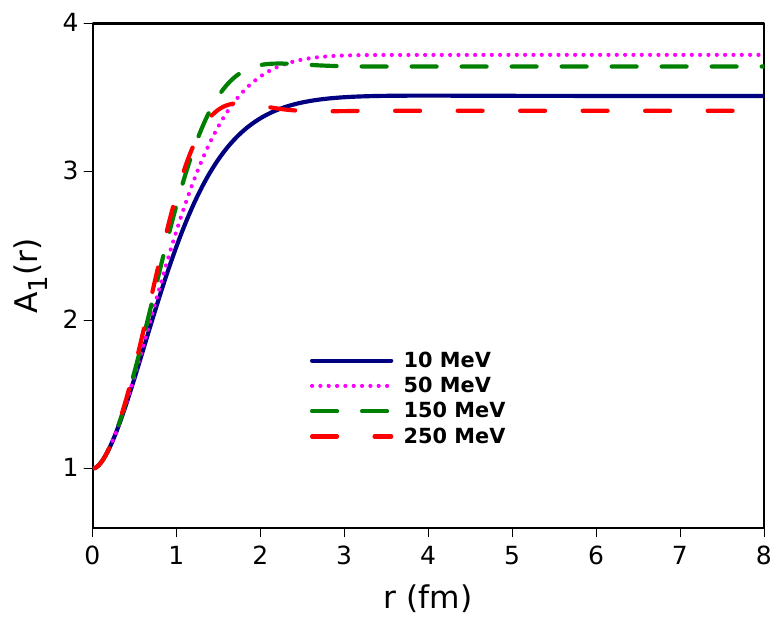}
    \includegraphics[width=0.55\linewidth]{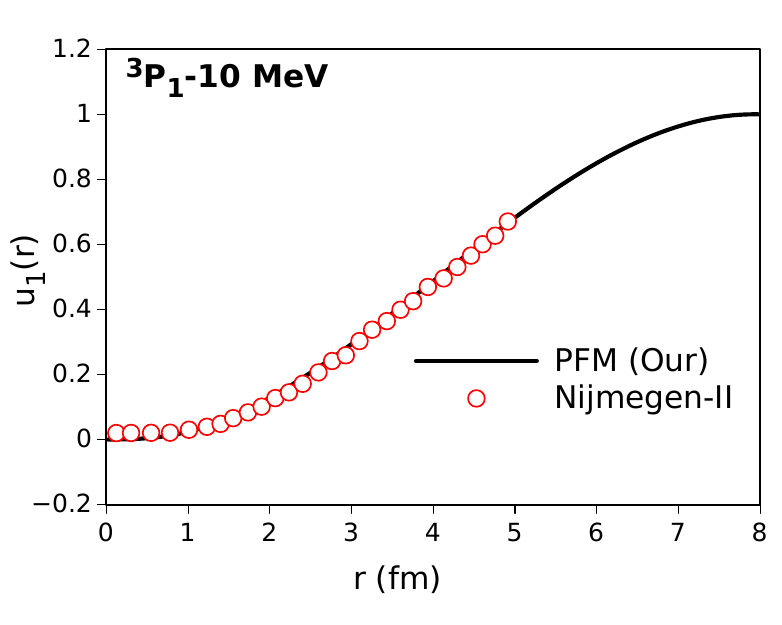}
    \includegraphics[width=0.55\linewidth]{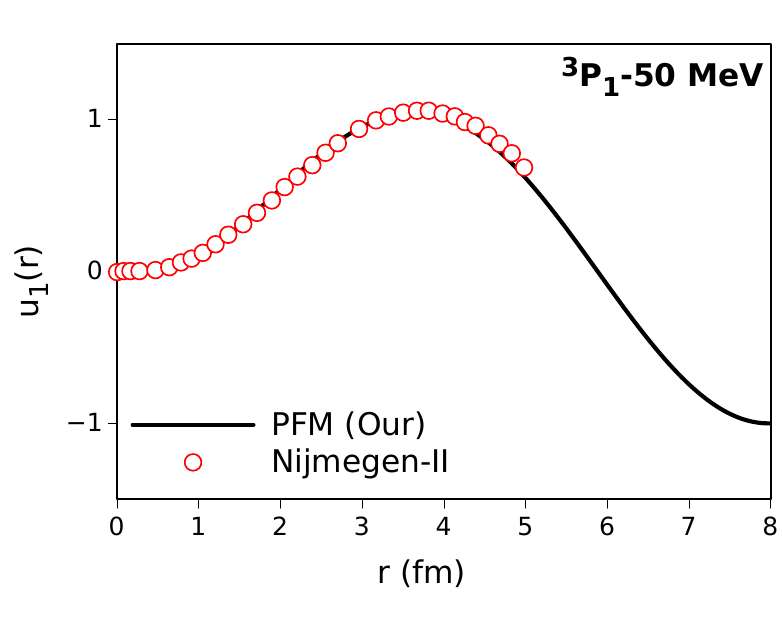}
    \includegraphics[width=0.55\linewidth]{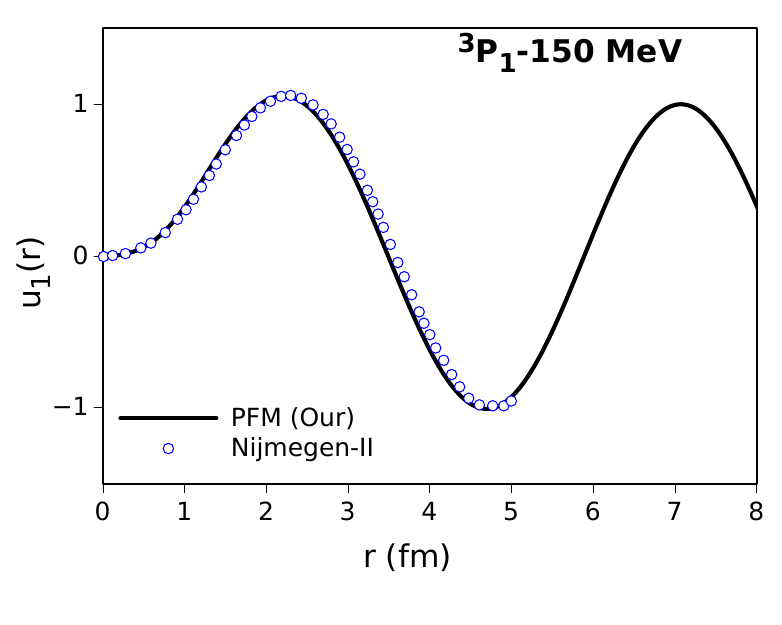}
    \includegraphics[width=0.55\linewidth]{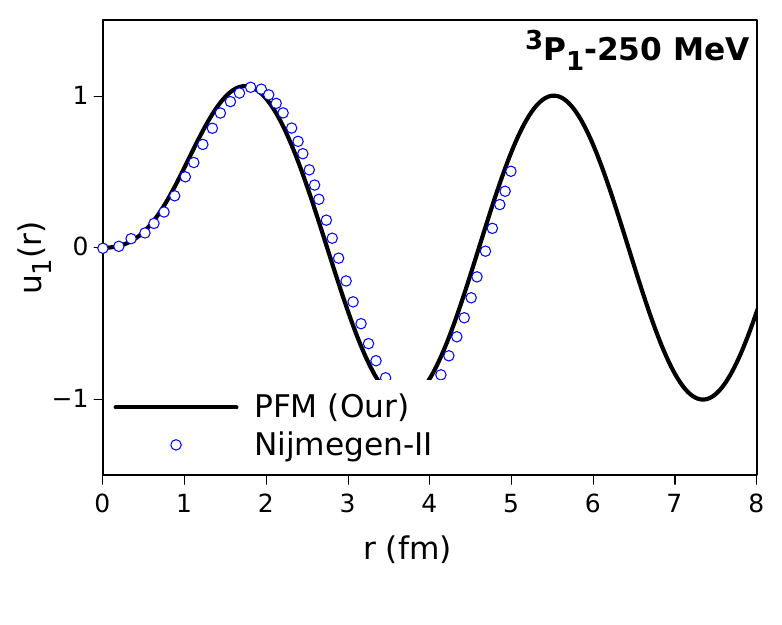}
    \caption{
Phase function, amplitude function, and radial wavefunctions for the $^{3}P_{1}$ channel.
Panel (a) shows the variation of the phase shift $\delta_{0}(r)$, panel (b) the amplitude
function $A_{0}(r)$, and panels (c--f) the corresponding wavefunctions $u_{0}(r)$ at
$E_{\text{lab}} = 10,\; 50,\; 150,$ and $250$~MeV. Results obtained using PFM are compared
with Nijmegen-II calculations.
}

    \label{fig:3D2}
\end{figure} 

 \begin{figure}
      \includegraphics[width=0.55\linewidth]{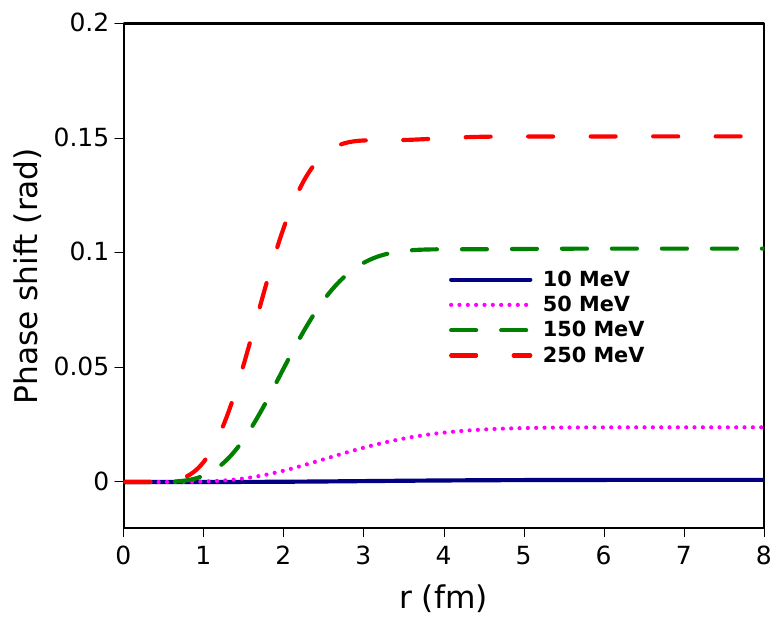}
      \includegraphics[width=0.55\linewidth]{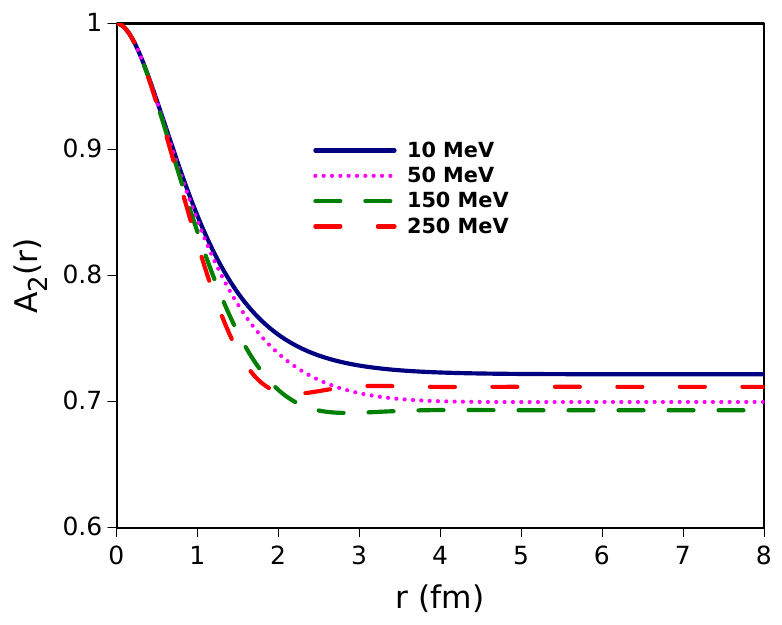}
    \includegraphics[width=0.55\linewidth]{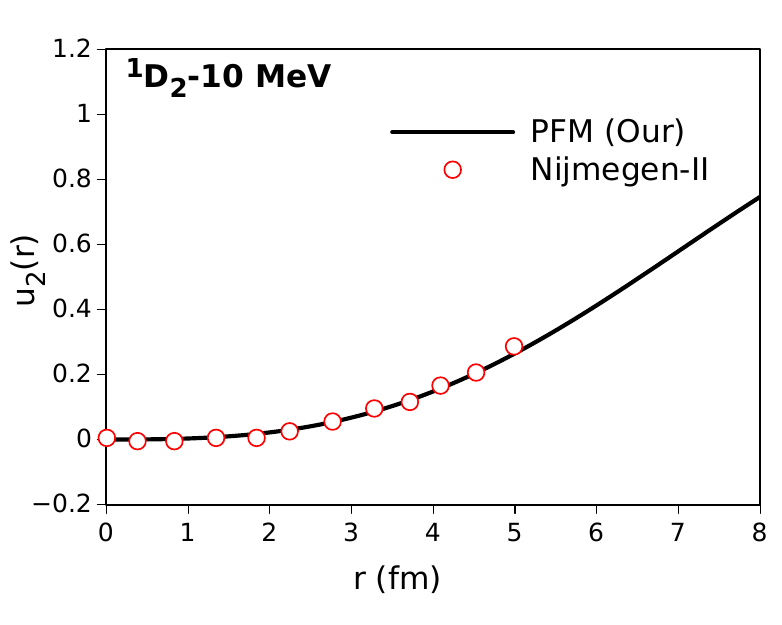}
    \includegraphics[width=0.55\linewidth]{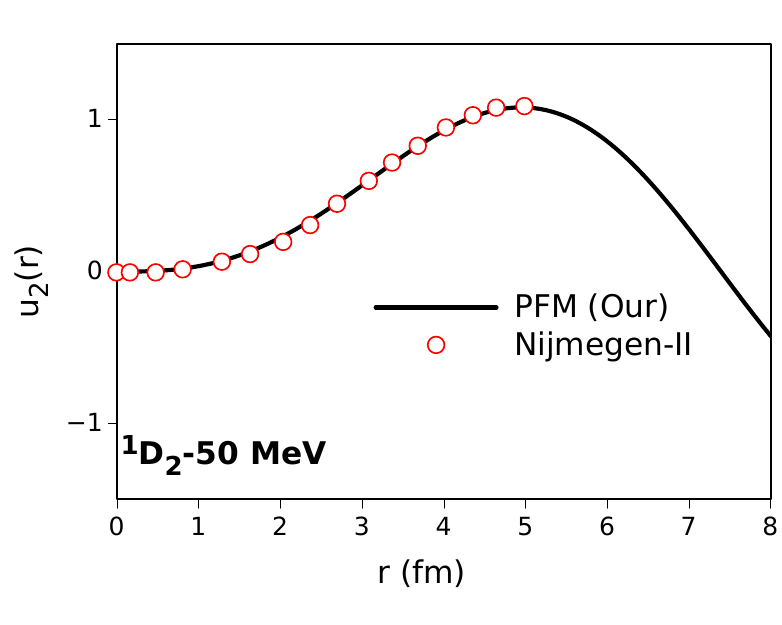}
    \includegraphics[width=0.55\linewidth]{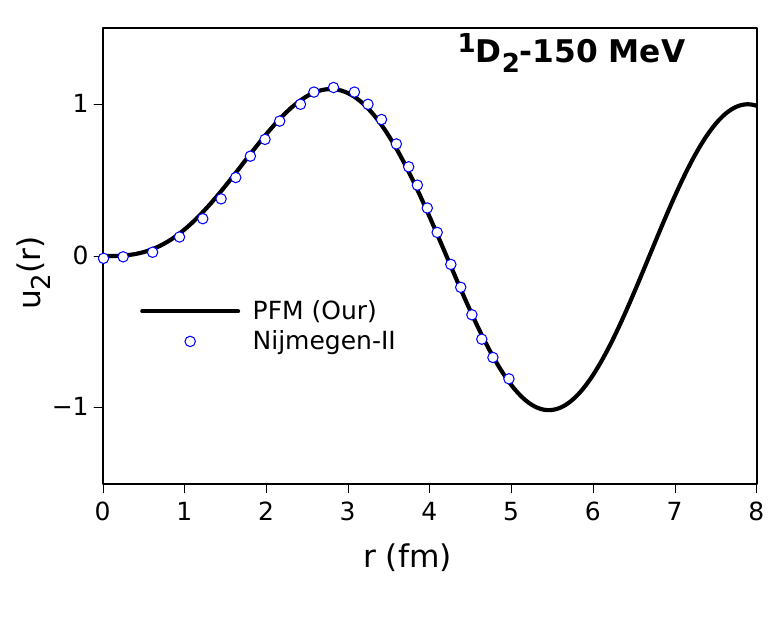}
    \includegraphics[width=0.55\linewidth]{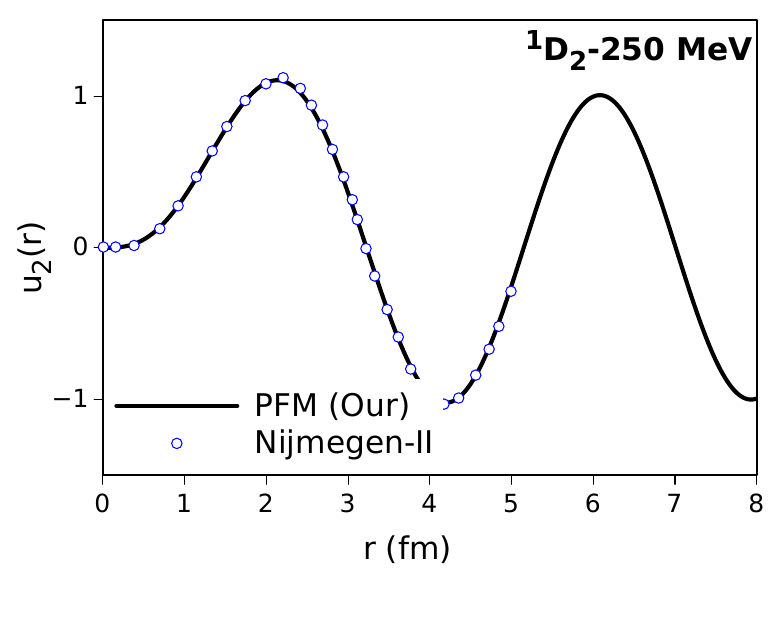}
    \caption{
Phase function, amplitude function, and radial wavefunctions for the $^{1}D_{2}$ channel.
Panel (a) shows the variation of the phase shift $\delta_{0}(r)$, panel (b) the amplitude
function $A_{0}(r)$, and panels (c--f) the corresponding wavefunctions $u_{0}(r)$ at
$E_{\text{lab}} = 10,\; 50,\; 150,$ and $250$~MeV. Results obtained using PFM are compared
with Nijmegen-II calculations.
}

    \label{fig:3D2}
\end{figure} 

 \begin{figure}
    \includegraphics[width=0.55\linewidth]{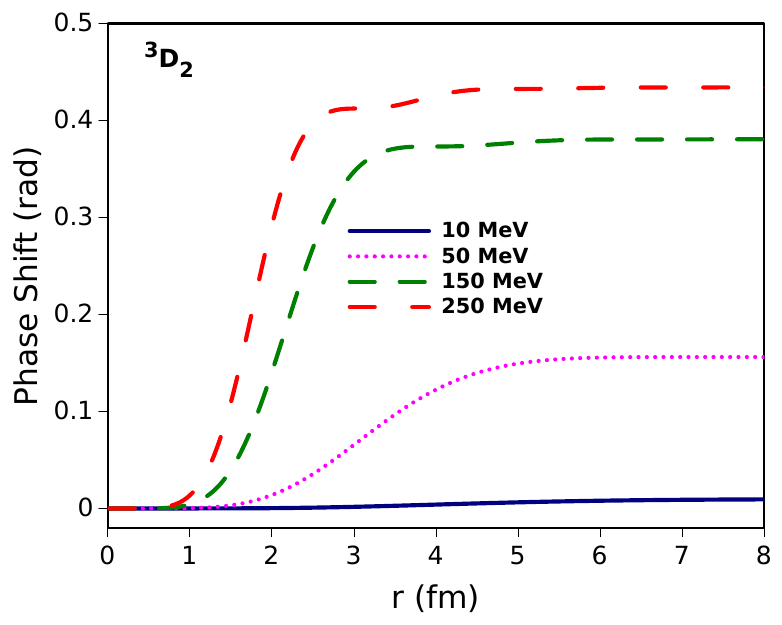}
    \includegraphics[width=0.55\linewidth]{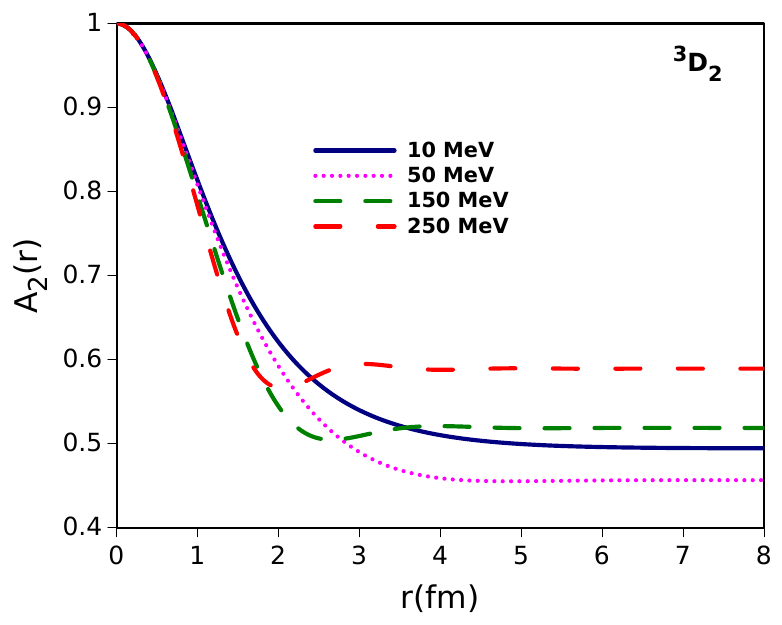}
    \includegraphics[width=0.55\linewidth]{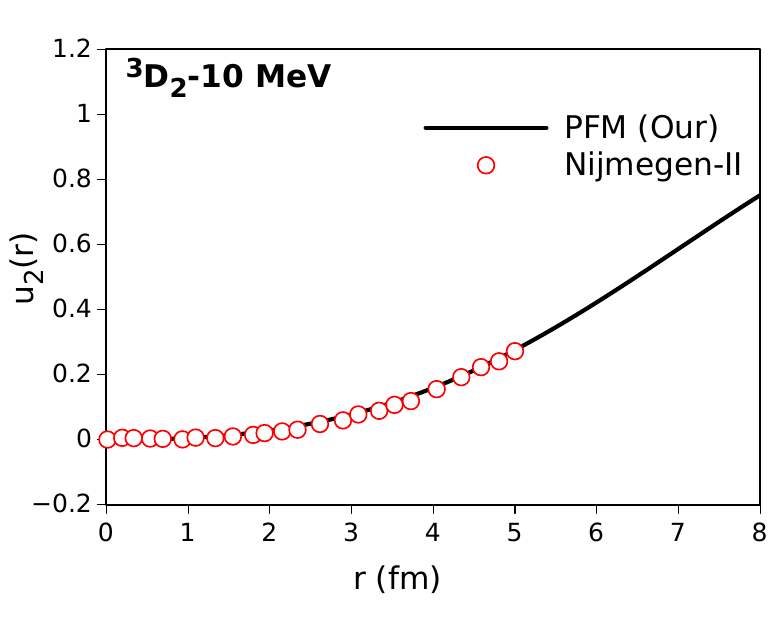}
    \includegraphics[width=0.55\linewidth]{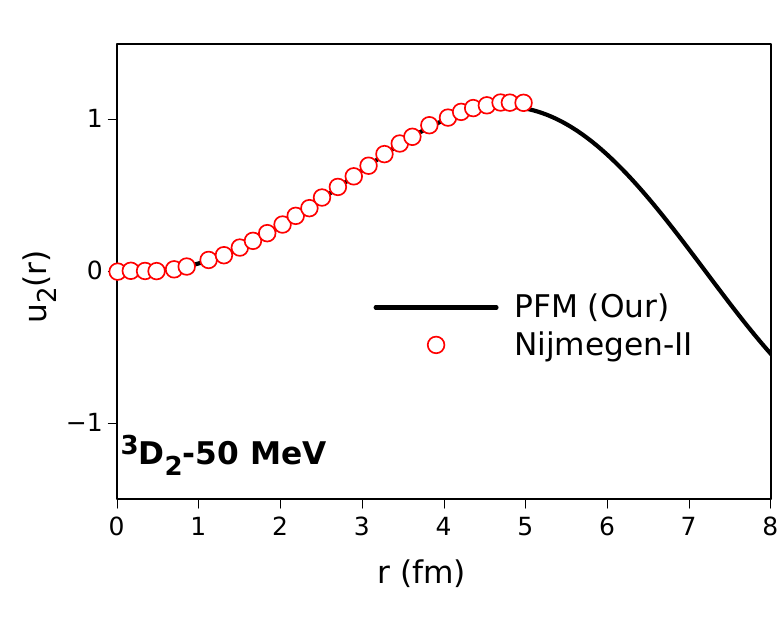}
    \includegraphics[width=0.55\linewidth]{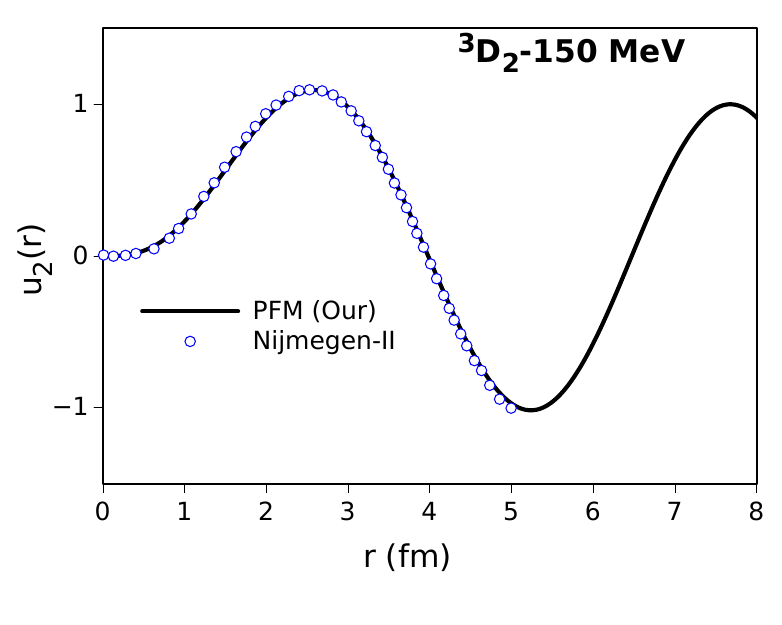}
    \includegraphics[width=0.55\linewidth]{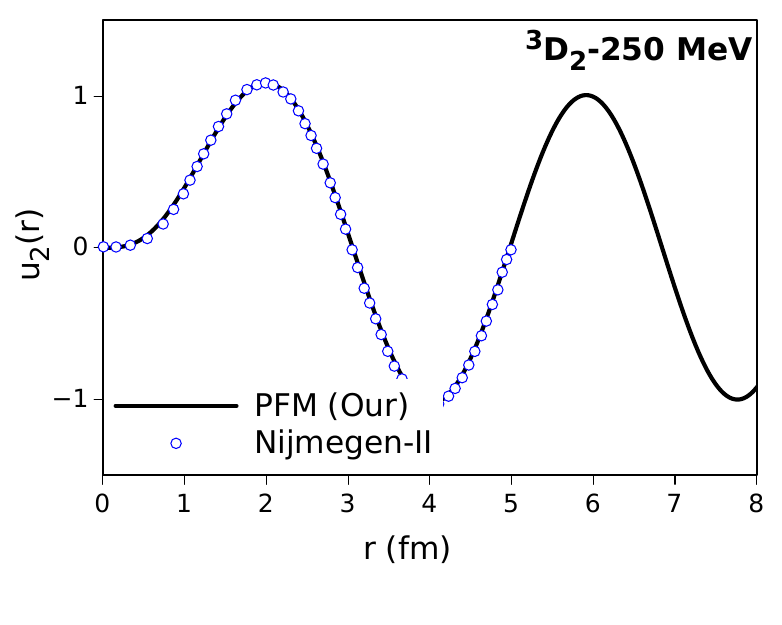}
    \caption{
Phase function, amplitude function, and radial wavefunctions for the $^{3}D_{2}$ channel.
Panel (a) shows the variation of the phase shift $\delta_{0}(r)$, panel (b) the amplitude
function $A_{0}(r)$, and panels (c--f) the corresponding wavefunctions $u_{0}(r)$ at
$E_{\text{lab}} = 10,\; 50,\; 150,$ and $250$~MeV. Results obtained using PFM are compared
with Nijmegen-II calculations.
}

    \label{fig:3D2}
\end{figure}

\subsection{Discussion}
\subsubsection{$^{1}S_{0}$ State}

The results for the $^{1}S_{0}$ channel are shown in Fig.~3, where the radial phase
shift, $\delta(r)$, amplitude function $A(r)$, and corresponding wavefunctions $u_0(r)$ are presented for
laboratory energies $E_{\text{lab}} = 10$, 50, 150, and 250~MeV. Owing to the absence
of a centrifugal barrier, the $S$-wave channel is particularly sensitive to the
short-range structure of the interaction and therefore provides a stringent test
of the inverse potential and the Phase Function Method.

The radial phase function $\delta_{0}(r)$, displayed in Fig.~3(a), exhibits a clear
energy-dependent behavior. At low energies, the phase shift accumulates gradually
with distance, reflecting the dominance of the attractive region of the
interaction. As the energy increases, the influence of the short-range repulsive
core becomes more pronounced, leading to a reduction and partial reversal of the
phase accumulation at small $r$. This behavior is fully consistent with the PFM
formalism, where repulsive regions contribute negatively and attractive regions
positively to the total phase shift.

The corresponding amplitude functions $A_{0}(r)$, shown in Fig.~3(b), vary smoothly
over the interaction range for all energies considered. The amplitude exhibits a
stronger radial modulation at higher energies, indicating increased sensitivity to
the inner region of the potential. The absence of numerical instabilities or
unphysical oscillations confirms the robustness of the first-order amplitude
equation and the reliability of the inverse Morse potential for the $^{1}S_{0}$
state.

Figures~3(c)--(f) present the radial wavefunctions $u_{0}(r)$ at
$E_{\text{lab}} = 10$, 50, 150, and 250~MeV, respectively, along with comparisons to
the Nijmegen-II interaction. At the lowest energy, noticeable differences in the
short-range behavior are observed, reflecting the sensitivity of low-energy
$S$-wave scattering to details of the interaction. With increasing energy, the
wavefunctions exhibit enhanced oscillatory behavior and show improved agreement
with the Nijmegen-II results over the entire radial range. The correct asymptotic behavior
and smooth matching across the interaction region further validate the present
approach.

Overall, the consistent behavior of the phase shift, amplitude function, and
wavefunctions for the $^{1}S_{0}$ channel demonstrates that the Phase Function
Method, combined with inverse Morse potentials, provides a reliable and physically
transparent description of $S$-wave neutron--proton scattering.
\subsubsection{$^{1}P_{1}$ State}

The results for the $^{1}P_{1}$ partial wave are presented in Fig.~4, where the
radial phase shift $\delta(r)$, amplitude function $A(r)$, and corresponding wavefunctions $u_1(r)$ are
displayed for laboratory energies $E_{\text{lab}} = 10$, 50, 150, and 250~MeV.
The inverse Morse potential for this channel is found to be purely repulsive in
nature (Figure 2 (dotted black)), which strongly influences the behavior of all scattering observables.

The variation of the phase function $\delta_{1}(r)$, shown in Fig.~4(a), exhibits
negative values over the entire interaction range for all energies. This behavior is a direct consequence of the repulsive character of the interaction, in full
accordance with the Phase Function Method, where a positive potential leads to a
negative phase accumulation. At lower energies, the magnitude of the phase shift
remains small, indicating that the scattering particles predominantly probe the
outer region of the interaction. As the energy increases, the particles penetrate
closer to the repulsive core, resulting in a more pronounced negative phase shift.

The corresponding amplitude functions $A_{1}(r)$, depicted in Fig.~4(b), show a
smooth and monotonic radial variation without any anomalous oscillations. The
energy dependence of the amplitude reflects the increasing influence of the
repulsive interaction at higher energies. The stability of $A_{1}(r)$ confirms the
numerical robustness of the PFM-based first-order formulation and validates the
quality of the inverse potential employed for this channel.

Figures~4(c)--(f) present the radial wavefunctions $u_{1}(r)$ at different
energies, along with comparisons to the Nijmegen-II interaction. At low energy
($E_{\text{lab}} = 10$~MeV), the wavefunction exhibits a suppressed amplitude at
short distances, characteristic of a repulsive $P$-wave interaction. With
increasing energy, the wavefunctions develop enhanced oscillatory behavior and
show improved agreement with the Nijmegen-II results, particularly beyond the core
region. The absence of spurious nodes and the smooth matching to asymptotic
behavior further demonstrate the physical reliability of the constructed
wavefunctions.

Overall, the consistent behavior of the phase shift, amplitude function, and
wavefunctions for the $^{1}P_{1}$ state confirms that the Phase Function Method,
combined with inverse Morse potentials, provides an accurate and transparent
description of repulsive $P$-wave neutron--proton scattering.

\subsubsection{$^{3}P_{0}$ and $^{3}P_{1}$ States}
The results for the $^{3}P_{0}$ and $^{3}P_{1}$ partial waves are shown in
Figs.~5 and 6, respectively, where the radial phase shifts, amplitude functions,
and corresponding wavefunctions are presented for
$E_{\text{lab}} = 10$, 50, 150, and 250~MeV. Although both channels correspond to
$P$-wave scattering, their interaction characteristics differ significantly and
are clearly reflected in the PFM observables.

For the $^{3}P_{0}$ state, the inverse potential contains both attractive and
repulsive components. As a result, the radial phase shift exhibits a transition
from negative to positive values with increasing distance, indicating the
interplay between short-range repulsion and intermediate-range attraction.
The associated amplitude function shows a pronounced energy dependence, while
the resulting wavefunctions display good agreement with Nijmegen-II calculations,
particularly at intermediate and higher energies.

In contrast, the $^{3}P_{1}$ channel is dominated by a repulsive interaction.
Consequently, the phase shift remains negative over the entire interaction range
for all energies considered, consistent with the expectations of the Phase
Function Method. The amplitude functions vary smoothly with distance, and the
wavefunctions exhibit the characteristic suppression at short distances typical
of repulsive $P$-wave interactions. Agreement with Nijmegen-II improves as the energy
increases, reflecting the reduced sensitivity to the detailed shape of the
short-range potential.

Overall, the distinct behaviors observed in the $^{3}P_{0}$ and $^{3}P_{1}$
channels demonstrate the ability of the Phase Function Method combined with
inverse Morse potentials to reliably capture both attractive--repulsive
interplay and purely repulsive dynamics in triplet $P$-wave neutron--proton
scattering.

\subsubsection{$^{1}D_{2}$ and $^{3}D_{2}$ States}

The results for the $^{1}D_{2}$ and $^{3}D_{2}$ partial waves are shown in
Figs.~7 and 8, respectively, where the radial phase shifts, amplitude functions,
and wavefunctions are presented for
$E_{\text{lab}} = 10$, 50, 150, and 250~MeV. Owing to the presence of the
centrifugal barrier, $D$-wave scattering is less sensitive to the short-range
core and predominantly probes the intermediate-range part of the interaction.

For the $^{1}D_{2}$ state, the inverse potential is purely attractive, leading to
a positive phase accumulation over the entire interaction range. The radial
phase shift increases smoothly with distance for all energies, in agreement
with the expectations of the Phase Function Method. The corresponding amplitude
functions exhibit weak energy dependence and remain stable throughout the
interaction region. The resulting wavefunctions show very good agreement with
Nijmegen-II calculations, indicating that the present inverse potential reliably
captures the dominant physics of this channel.

A similar behavior is observed for the $^{3}D_{2}$ state, where the interaction
is also predominantly attractive. The phase shift remains positive for all
energies, while the amplitude function varies smoothly with distance. The
wavefunctions display regular oscillatory behavior and compare well with Nijmegen-II
results, particularly at intermediate and higher energies. Minor differences
observed at short distances can be attributed to the simplified form of the
Morse potential and the reduced sensitivity of $D$-waves to short-range details.

Overall, the consistent behavior of the phase shifts, amplitude functions, and
wavefunctions in the $^{1}D_{2}$ and $^{3}D_{2}$ channels further demonstrates
the effectiveness of the Phase Function Method combined with inverse Morse
potentials in describing higher partial-wave neutron--proton scattering.

The present analysis is restricted to uncoupled neutron--proton scattering
channels, for which the Phase Function Method can be formulated within a
single-channel framework. Coupled-channel effects, which are known to play a
significant role in states such as $^{3}S_{1}$--$^{3}D_{1}$ due to tensor
interactions, are not included in the current work. The extension of the present
inverse-potential and PFM-based formalism to coupled channels requires the
treatment of matrix-valued phase and amplitude functions and is therefore
computationally more involved. Such an extension constitutes a natural
continuation of the present study.

\section{Conclusion} 
In this work, the Phase Function Method (PFM) has been successfully employed to construct distance-dependent phase functions, amplitude functions, and explicit radial wavefunctions for neutron--proton scattering in various uncoupled partial waves. The interaction parameters were taken from inverse Morse potentials previously obtained by fitting high-quality experimental phase-shift data, thereby providing a consistent and physically motivated framework.

The radial evolution of the phase shift $\delta(r)$ clearly reflects the underlying nature of the interaction, with repulsive regions leading to negative phase accumulation and attractive regions contributing positively, in full agreement with the PFM formalism. The corresponding amplitude functions exhibit smooth and stable behavior over the entire interaction range, confirming the numerical
robustness of the approach.

The obtained wavefunctions show very good agreement with those generated using high-precision nucleon--nucleon interactions such as Nijmegen-II, NIJM-II, and Av18, particularly at intermediate and higher energies. Small phase differences observed in certain channels can be attributed to the simplified form of the Morse potential
rather than limitations of the PFM itself.

Overall, the present study establishes PFM as an efficient and transparent computational tool for obtaining realistic scattering wavefunctions directly from inverse potentials. The methodology can be readily extended to other two-body scattering systems, including $n\alpha$, $p\alpha$, and heavier nuclear systems, making it a promising approach for future scattering and structure
studies. 

Since coupled channel calculations are missing the calculations its be the next part of paper.

\section{Acknowledgements}
I am grateful to Prof. Calogero for his work on potential scattering.
\section*{Appendix} 
\subsection*{Amplitude Function Equations}
\vspace{.1cm}
for $\ell=0$,$ \ell=1$, $\ell=2$
\vspace{.1cm}
\begin{align}
A_{0}^{\prime} &= -\frac{A_{0} V(r)}{k\left(\frac{\hbar^{2}}{2\mu}\right)} \left[\cos \delta_{0} \cdot \sin(kr) - \sin \delta_{0} \cdot (-\cos(kr))\right] \nonumber \\
&\quad \times \left[\sin \delta_{0} \cdot \sin(kr) + \cos \delta_{0} \cdot (-\cos(kr))\right] \\
\\
A_{1}^{\prime} &= -\frac{A_{1} V(r)}{k\left(\frac{\hbar^{2}}{2\mu}\right)} \left[ \cos \delta_{1} \left( \frac{\sin (kr)}{(kr)}-\cos (kr) \right) - \sin \delta_{1} \left( -\frac{\cos (kr)}{(kr)}-\sin (kr) \right) \right] \nonumber \\
&\quad \times \left[ \sin \delta_{1} \left( \frac{\sin (kr)}{(kr)}-\cos (kr) \right) + \cos \delta_{1} \left( -\frac{\cos (kr)}{(kr)}-\sin (kr) \right) \right]
\\
A_{2}^{\prime} &= -\frac{A_{2} V(r)}{k\left(\frac{\hbar^{2}}{2\mu}\right)} \biggl[\cos \delta_{2} \left(\left(\frac{3}{(k r)^{2}}-1\right) \sin (kr)-\frac{3}{(k r)} \cos (kr)\right) \nonumber \\
&\quad \times \left(\sin \delta_{2} \left(\left(\frac{3}{(k r)^{2}}-1\right) \sin (kr)-\frac{3}{(k r)} \cos (kr)\right) + \cos \delta_{2} \left(\left(-\frac{3}{(k r)^{2}}+1\right) \cos (kr) \right. \right. \nonumber \\
&\quad \left. \left. -\frac{3}{(k r)} \sin (kr)\right)\right]
\end{align}
\subsection*{Wavefunction Equations}
\vspace{.1cm}
for $\ell=0$, $\ell=1$, $\ell=2$
\vspace{.1cm}
\begin{flalign}
    &u_{0}(r) = A_{0}(r) \left[\cos \delta_{0}(r) \cdot \sin(kr) - \sin \delta_{0}(r) \cdot \cos(kr)\right] & \\
    &u_{1}(r) = A_{1}(r) \left[\cos \delta_{1}(r) \left(\frac{\sin (kr)}{(kr)}-\cos (kr)\right) - \sin \delta_{1}(r) \left(-\frac{\cos (kr)}{(kr)}-\sin (kr)\right)\right] & \\
    &u_{2}(r) = A_{2}(r) \bigg[\cos \delta_{2}(r) \bigg(\left(\frac{3}{(k r)^{2}}-1\right) \sin (kr)-\frac{3}{(k r)} \cos (kr)\bigg) \nonumber \\
    &\qquad - \sin \delta_{2}(r) \bigg(-\left(-\frac{3}{(k r)^{2}}+1\right) \cos (kr)-\frac{3}{(k r)} \sin (kr)\bigg)\bigg] &
\end{flalign}

\end{document}